%% file: infert.tex
\pdfoutput=1

\documentclass[10pt,letter,twocolumn]{article}
\usepackage{f1000_styles}

\usepackage[super,comma,sort]{natbib}
\usepackage[colorlinks=true,urlcolor=blue]{hyperref}

\usepackage{amsmath}
\input infert-def

\def\citeyear{\citep}
\def\autocite{\citep}
\def\textcite{\citet}

\begin{document}

\title{Puzzles in modern biology. I. Male sterility, failure reveals design}
\author[1]{Steven A.~Frank}
\affil[1]{Department of Ecology and Evolutionary Biology, University of California, Irvine, CA 92697--2525 USA, safrank@uci.edu}

\maketitle
\thispagestyle{fancy}

\parskip=4pt

\begin{abstract}

Many human males produce dysfunctional sperm. Various plants frequently abort pollen. Hybrid matings often produce sterile males. Widespread male sterility is puzzling. Natural selection prunes reproductive failure. Puzzling failure implies something that we do not understand about how organisms are designed. Solving the puzzle reveals the hidden processes of design. 

\end{abstract}


\vskip0.5in
\noterule
Preprint of published version: Frank, S. A. 2016. Puzzles in modern biology. I. Male sterility, failure reveals design. F1000Research 5:2088,  \hfil\break \href{http://dx.doi.org/10.12688/f1000research.9567.1}{doi:10.12688/f1000research.9567.1}. \break Published under a Creative Commons \href{https://creativecommons.org/licenses/by/4.0/}{CC BY 4.0} license.
\smallskip
\noterule

\clearpage
\subsection*{Introduction}
What do organisms do poorly that they should do well? Answers often lead to profound insight.

One thing organisms should do well is reproduce. Sterility is total biological failure. Natural selection prunes failure. Widespread sterility would be puzzling.

Yet, many human males produce incapable sperm. Some plants have up to 50\% of individuals abort their pollen. Matings between recently diverged species often bear sterile males.

Causes of sterility vary. Each cause is a separate topic. But separating topics hides the deeper unity of insight. \textit{Through failure we understand design}.

I briefly review examples of male sterility. These individual puzzles emerged haphazardly, rather than by systematic study of failure. From these examples, I return to my theme. Every apparent failure poses an important puzzle. We must seek failure, measure it, document its correlates, and analyze its causes.

\subsection*{Sperm dysfunction}
Sperm dysfunction poses our first puzzle. Roughly 5\% of human males fail to make good sperm\autocite{krausz11male}. In other animals, studies mention cases of male sterility\autocite{jainudeen00reproductive}. These haphazard observations, based on limited data, provide only a vague hint. Puzzles often appear in this shadowy way. Let us follow the shadow. If sterility is higher than expected, what might explain the excess?

A quirk of genetics predisposes males to failure\autocite{frank96mitochondria}. A male inherits his mother's mitochondria, but does not pass mitochondrial genes to his progeny. Natural selection cannot act on male-specific mitochondrial effects, because males do not transmit mitochondria. Mitochondrial mutations that reduce male fertility may increase by chance. 

Studies motivated by this theory have found mitochondrial mutations that reduce male fertility\autocite{dowling08evolutionary,patel16a-mitochondrial}. Those studies have also found other genes, inherited through both parents, that compensate for the mitochondrial defects. A male carrying the mitochondrial mutation and the compensatory genes have restored fertility. 

Compensation arises from the pathways of inheritance\autocite{frank96mitochondria}. A gene inherited through both parents suffers reduced transmission when coupled with mitochondrially induced sperm defects. If a biparentally inherited gene compensates for the defects, the compensatory gene increases its own transmission.

Compensation plays a key role in failure and design. In this case, biparentally inherited genes compensate for male-sterile mitochondrial mutations. Individuals appear to be nearly normal when carrying both the defect and the compensation. Failure occurs only when there is a mismatch in the coadapted defect-compensation interaction. 

At present, only a few studies support these ideas about mitochondrial transmission and genetic coadaptation. Future studies may provide further support. Or it may turn out that other processes explain much of the observed failure. The puzzle remains unsolved. But we have a clue that points to disruption of coadapted gene complexes.

I now turn to other examples of sterility and coadaptation. These additional puzzles provide a broader perspective on the nature of failure and design. Perhaps some of the insights from solving these additional puzzles will eventually lead back to better explanations for the apparently high levels of sperm dysfunction.

\subsection*{Aborted pollen}
Aborted pollen poses our next puzzle. Most flowering plants are hermaphrodites. Each individual produces both ovules and pollen. Ovules correspond to female function. Pollen correspond to male function. 

Some hermaphroditic individuals abort their pollen. Rare male sterility within a population would not be surprising. Most traits fail in a few individuals. However, Darwin\autocite{darwin77the-different} noted puzzling, widespread male sterility in many different plant species. Several populations have more than 10\% male sterility, sometimes approaching 50\%. 

Two explanations of this apparent failure alter our perspective of design. First, male sterility prevents self-fertilization of ovules. If progeny suffer when inbred, the prevention of self-fertilization can be advantageous\autocite{charlesworth78a-model}. The gain from outbred ovules can outweigh the loss of pollen production by male sterility. If so, then male sterility is a beneficial design rather than a failure. 

Some studies support the outbreeding benefit of male sterility\autocite{weller05selfing}. The initial surprise of a high failure rate has become a deeper insight into organismal design. However, many populations have high frequencies of male sterility that cannot be explained by avoiding self-fertilization\autocite{charlesworth81a-further}.

This puzzling excess of male sterility led to a second explanation\autocite{lewis41male,charlesworth81a-further,frank89the-evolutionary}. Mitochondria transmit only through ovules, the female lineage. Pollen do not transmit mitochondria. A mitochondrial mutation would gain a benefit by aborting pollen and reallocating the saved energy to produce more successful ovules. The pollen-aborting mitochondria increase their transmission and can spread rapidly in populations. Mitochondrial mutations that abort pollen and enhance ovule success have been found in many species that previously had an unexplained excess of male sterility\autocite{chase07cytoplasmic}. 

In plants, the vast majority of genes transmit biparentally, through both pollen and ovules. Those biparental genes typically suffer reduced transmission when in a plant that aborts its pollen. In most cases, when a mitochondrial mutation exists that aborts pollen, there also exist biparentally inherited genes that can restore pollen fertility\autocite{chase07cytoplasmic}. 

An apparently normal hermaphrodite with full pollen fertility may often carry two opposing components. First, a mitochondrial mutation that, unblocked, causes male sterility. Second, a biparentally inherited gene that blocks the action of the mitochondrial mutation. Once again, ``normal'' function arises by compensatory coadaptation.  

Crosses between different species support this idea of coadaptation\autocite{laser72anatomy}. The parental populations may be almost entirely free of male sterility. Yet the hybrid progeny may express high frequencies of aborted pollen, which is male sterility. 

In some cases, the hybrid male sterility arises by breaking up the coadapted complexes within each species. A hybrid progeny may carry the male sterile mitochondrial mutation of its ``mother'' but fail to inherit from its ``father'' the associated restorers of male fertility. 

\subsection*{Hybrid male sterility}
A different kind of hybrid male sterility poses our final puzzle. Matings between species often produce defective progeny. Male sterility is one of the most common hybrid defects\autocite{coyne04speciation,schilthuizen11haldanes,delph16haldanes}. The puzzle concerns why male fertility should be particularly prone to failure. What does that failure reveal about design? 

The previous puzzles raised two potential causes of hybrid male sterility. In each case, mitochondrial mutations disrupt male fertility. In response, biparentally inherited genes evolve to repress the disruption. The coadapted repressor genes restore male fertility. Hybrid matings cause mismatch of coadapted genes, leading to male sterility.

Not all cases of hybrid male sterility arise from the breakdown of coadaptation between mitochondrial and biparental genes. What other aspects of design might lead to the observed widespread tendency for failure in hybrid males?

Hybrid defects typically arise from mismatch of coadapted genes. For male fertility, how does coadaptation evolve? What causes divergence between populations in their coadapted complexes?

Our previous puzzles suggest how we might think about these broader questions. In the first puzzle of sperm dysfunction, males do not transmit mitochondrial mutations. Any mutation that influences only males has no consequence for transmission. This neutrality means that male sterile mutations can increase unopposed. But such mutations are not directly favored. 

Mitochondrial mutations that disrupt male fertility accumulate slowly, by chance. Different populations accumulate different mutations. Each distinct mitochondrial mutation associates with distinct compensatory mechanisms of biparental genes. 

As populations diverge, they will come to have different coadapted complexes of mitochondrial and biparental genes. The neutral accumulation of mitochondrial mutations causes slow but continual divergence of coadapted complexes. 

In the second puzzle of pollen abortion, mitochondrial mutations also disrupt male fertility. However, the hermaphroditic system means that the decline in male fertility often associates with a rise in female fertility. A mitochondrial mutation that causes male sterility gains a transmission advantage through its increased female fertility. This benefit can drive rapid spread of mitochondrial mutations. 

In this case, mitochondrial mutations that disrupt male fertility spread rapidly. In response, the associated biparental restorers of fertility will likely evolve rapidly. The conflict between different components of the genome causes populations to diverge rapidly in their coadapted gene complexes. 

The two cases set endpoints for the range of processes that cause hybrid male sterility. On the one hand, ubiquitous neutral divergence occurs widely but relatively slowly. On the other hand, powerful conflict between components of the genome drives rapid divergence, but may arise relatively rarely. 

The continuum ranges from ubiquitous and slow processes to less common and fast processes. Many processes fall along this continuum. The relative roles of these different process for hybrid male sterility remain controversial. 

For a particular observation of male sterility, we may not know the particular associated process. However, my point is that we should pay attention to the observed failure. Through the study of failure, we gain a window onto the normally hidden underlying processes of design. 

For hybrid male sterility, I describe one example of a ubiquitous but relatively slow process and one example of a rarer but relatively fast process. That contrast highlights a potentially important question about failure and design. Which tends to be more important, slow processes intrinsic to ubiquitous aspects of genetics or fast processes intrinsic to specific aspects of conflict?

The example of relatively slow ubiquity concerns the spread of beneficial mutations on the sex chromosomes\autocite{charlesworth87the-relative}. In animals with separate sexes, males often carry a pair of different sex chromosome types, XY, whereas females carry a pair of the same chromosome type, XX. 

Consider a new mutation on the X chromosome that benefits only females. Because the X chromosome in females occurs in two copies, the new mutation on one X may be masked by the expression of the original gene carried on the other X. That masking effect can greatly reduce the rate at which beneficial mutations can spread.

Now consider a new mutation on the X chromosome that benefits only males. Because the X chromosome in males occurs in only one copy, the new mutation is not masked by a different copy of the gene on another chromosome. The beneficial mutation is expressed immediately and can spread.

This asymmetry leads to faster evolution of male-specific effects on the X chromosome. As those X-linked effects evolve, other components of the genome may coadapt. Populations will diverge in coadapted gene complexes between X-linked genes and genes in other parts of the genome.

When diverged populations hybridize, the faster-evolving X-linked male effects may be particularly susceptible to the mismatch of coadapted complexes. Some of those mismatches may reduce male fertility. 

In this case, the divergence of coadapted complexes happens in a relatively passive way. The structure of the genetic system creates an asymmetric sieve. That sieve tends to enhance male-specific effects more strongly than female-specific effects. Coadaptation arises as a potentially weak response to general aspects of change rather than a strong response to a direct challenge.

A contrasting example of relatively fast specificity concerns a conflict between different components of the genome. That conflict creates a direct and powerful pressure for change and coadaptation.

Once again, we begin with the XY chromosome pairing in males. A male transmits his X chromosome to daughters and his Y chromosome to sons. An X chromosome gains an advantage by increasing a male's number of daughters. For example, an X can encode a mechanism that kills off Y bearing sperm. The male's remaining sperm bear the X, so he produces all daughters. 

This drive of X against Y occurs in some organisms\autocite{meiklejohn10genetic}. The driving X gains a transmission advantage and can spread rapidly within populations. A driving X favors genes on other chromosomes that repress the drive. The other genes suffer because the driving X biases the sex ratio toward disadvantageous over-production of daughters. The driving X may also carry deleterious side effects.

A driving X and coadapted repression of drive are powerful forces. Those powerful forces cause rapid change in populations and rapid divergence of coadapted complexes between populations\autocite{frank91divergence,hurst91causes}. 

In natural populations, some processes of divergence will be like the asymmetric XY sieve. Change accumulates by a relatively nonspecific process, ubiquitous but relatively slow. Other processes will be like the conflict of XY drive. Change accumulates by the strong process of transmission bias, narrowly specific but relatively fast. 

The importance of different processes must be determined by direct observation. My point concerns the value of focusing on failure. Tracing the observed failure of male sterility to its underlying cause will teach us much about the different processes that coadapt genomes and cause divergence between populations.

\subsection*{Conclusions}

These various puzzles bring us to a broader question. Why does the study of failure reveal underlying design? 

Roughly speaking, we tend to see organisms as reasonably well designed engineering solutions. But we do not know how the components have been put together, how the components interact, and how the components have been designed to respond to different challenges. How can we reverse engineer the design? 

The study of failure provides an important tool for reverse engineering. Why? Because when something complicated works, it is not easy to see how the components interact. Imagine that you have never seen a car. Someone gives you a car and asks you how it works. Look inside. There are many wires and connectors and components. What do they do? 

Try cutting a wire. The brakes fail. Through that failure, you know that the wire and the things connected by the wire have to do with braking. Failure reveals design.

\subsection*{Competing interests}
No competing interests were disclosed.

\subsection*{Grant information}
National Science Foundation grant DEB--1251035 supports my research.

{\small\bibliographystyle{unsrtnat}
\bibliography{infert}}

\end{document}

%% file: infert-def.tex
\newcount\BoxNum \BoxNum 1\relax
\makeatletter
\newcommand*{\boxlabel}[1]{%
  \protected@write \@auxout {}{\string \newlabel {box:#1}{{\the\BoxNum}}{}}%
  \advance\BoxNum 1\relax}
\makeatother

\newcommand*{\boldrule}{\hrule height 1.2pt}
\newcommand*{\noterule}{\medskip\boldrule\medskip}	